# Physics-based bias-dependent compact modeling of *1/f* noise in single- to few- layer 2D-FETs

Nikolaos Mavredakis*[a], Anibal Pacheco-Sanchez[a], Md Hasibul Alam[b], Anton Guimerà-Brunet[c, d]
Javier Martinez[c], Jose Antonio Garrido[e, f], Deji Akinwande[b] and David Jiménez[a]

*1/f* noise is a critical figure of merit for the performance of transistors and circuits. For two-dimensional devices (2D-FETs), and especially for applications in the *GHz* range where short-channel FETs are required, velocity saturation (VS) effect can result in the reduction of *1/f* noise at high longitudinal electric fields. A new physics-based compact model is for the first time introduced for single- to few layer 2D-FETs in this study, precisely validating *1/f* noise experiments for various types of devices. The proposed model mainly accounts for the measured *1/f* noise bias dependence as the latter is defined by different physical mechanisms. Thus, analytical expressions are derived, valid in all regions of operation in contrast to conventional approaches available in literature so far, accounting for carrier number fluctuation (*ΔN*), mobility fluctuation (*Δμ*) and contact resistance (*ΔR*) effects based on the underlying physics that rules these devices. *ΔN* mechanism due to trapping/detrapping together with an intense Coulomb scattering effect, dominates *1/f* noise from medium to strong accumulation region while *Δμ* is also demonstrated to modestly contribute in subthreshold region. *ΔR* can also be significant in very high carrier density. The VS induced reduction of *1/f* noise measurements at high electric fields, is also remarkably captured by the model. The physical validity of the model can also assist in extracting credible conclusions when conducting comparisons between experimental data from devices with different materials or dielectrics.

## Introduction

The discovery and first demonstration of graphene back in 2004[1] has been a landmark for the breakthrough of two-dimensional (2D) material's technology in nanoelectronics. The exceptional characteristics of graphene such as its high carrier mobility and saturation velocity, have led to the broad fabrication of graphene field-effect transistors (GFETs) and the design of analog/RF GFET-based circuits.[2] To the contrary, GFETs are inapplicable to digital applications due to a lack of bandgap. The latter has directed the scientific community to investigate other 2D materials such as transition metal dichalcogenides (TMDs) with unique electrical and optical properties,[3-5] for the design of next-generation ultra-scaled FETs. The aforesaid materials have been proven suitable for switching applications and logic circuits as they achieve a *1 to 2 eV* bandgap depending on their thickness. Hence, the specific 2D-FETs have obtained enormous consideration as they accomplish remarkable mobilities that can be superior to silicon while they present optimum electrostatic control of the gate over the channel due to their atomic thickness resulting in excellent scaling to short gate lengths.[6-7]

Numerous materials have been employed so far for the design and fabrication of 2D-FETs including $MoS_2$,[8] $MoSe_2$,[9] $WS_2$,[10] $MoTe_2$,[11] $ReS_2$[12] and black-phosphorous (BP)[13] and a significant bunch of applications has been developed to date by utilizing these devices, including low power digital and analog electronics,[14-16] optoelectronics,[17-18] nanophotonics[19] and chemical-biological sensors.[20-22] The efficiency of the aforementioned circuits can be degraded by the Low-frequency noise (LFN) of the intrinsic device. More specifically, LFN can restrict the sensitivity of sensors while it can be upconverted to unwelcome phase noise in oscillators and thus, deteriorate the quality of any wireless communication. Likewise, it can confine the level of signal that can be processed in digital applications due to high noise-to-signal ratio that is introduced in small devices. In addition, experimental LFN examination can reveal critical characteristics regarding various inherent mechanisms as surface trapping and Coulomb-induced charge fluctuations that can raise reliability issues and decay the quality of the device under test (DUT).[23-24]

LFN is generated due to three fundamental mechanisms in electronic devices. Firstly, the carrier number fluctuation effect as a result of trapping/detrapping of slow traps near the oxide interface, represented by a *ΔN* model.[25] In this effect, each carrier that gets trapped and then emitted back near the Fermi level, generates a random telegraph signal (RTS) which is transformed in a Lorentzian shape power spectral density (PSD) in frequency domain. The notable *1/frequency* (*f*) shape of LFN is recorded in long-gated devices due to the summation of an abundant number of the aforementioned Lorentzian spectra as the number of active traps is proportional to the device area, provided that the distribution of the time constants of the specific traps is uniform on the logarithmic axis. These time constants follow a non-radiative multiphonon model since a thermally activated process is considered.[25] The mobility fluctuation effect, associated with a *Δμ* model, is the second physical mechanism that can create LFN and is subject to the empirical Hooge formulation,[26] whereas LFN can also be produced in contacts apart from the channel due to series resistance, known as *ΔR* model.[27]

Several *1/f* noise modelling approaches in silicon devices refer only to *ΔN* model while they are simplified by considering a uniform channel leading to a direct $\sim(g_m/I_D)^2$ relation of the normalized with squared drain current, output noise $S_{ID}f/I_D^2$ where $g_m$ is the transconductance and $I_D$ the drain current of the device.[23, 28-30] This approximation is valid only in linear region and strong inversion of MOSFETs thus, a more complete physics-based model had been later introduced[27] (Section 6.3) which takes into account all the three effects (*ΔN*, *Δμ*, *ΔR*) as well as non-homogeneities in the channel present at higher drain voltage $V_{DS}$ regime. Coulomb scattering effect can also contribute to *ΔN* LFN through correlated mobility fluctuations (*ΔN CMF*) due to trapping processes. The latter mechanism has also been included in compact models.[23, 27, 30-32] The simplified model proposed in Ref. 32 is extended to saturation region for *ΔN CMF* (cf. equation 15), but does not end up with a compact formulation, which is required for including models in circuit simulators.

LFN in 2D-FETs has been thoroughly inspected focusing mainly on its experimental characterisation.[33-51] In more detail, LFN has been researched in $MoS_2$,[33-37, 39, 44, 46-47, 49-51] $MoSe_2$,[38] $WS_2$,[40] $MoTe_2$,[41-42] $ReS_2$[45, 48] and BP[44] FETs, respectively while single-,[33, 35-36, 39-40, 43, 46, 49, 51] bi-,[34, 36-37, 45, 49-50] and a few-[37-40, 44, 47-48, 51] layer (SL, BL, FL, respectively) devices have also been investigated as well as, liquid-gated[48] and paper-printing[50] 2D-FETs. Experimental random telegraph noise (RTN) has been recorded only at very low temperatures in 2D-FETs so far,[33] while it is usually the case in small devices, where the number of traps is small;[25] RTN is out of the scope of the present work. The preceding studies,[33-51] mostly report on ways to optimize *1/f* noise during fabrication of 2D-FETs due to the still premature stage of this technology. Hence, thermal annealing has been proposed as an efficient procedure to lower *1/f* noise[34, 36] while a surface passivation layer with a high-k dielectric ($Al_2O_3$[35, 41], $HfO_2$[42]) can also reduce it, as atmospheric contaminations such as residues, defects, water vapor etc. can be diminished. A hexagonal Boron Nitride (hBN) layer between the channel and the dielectric has been evidenced to improve *1/f* noise,[41, 48, 50] while the latter can also be decreased after the application of polymer electrolyte encapsulation.[46] *1/f* noise from the contacts can also contribute to total noise especially at high current regime[36-38] and vacuum annealing might be a decent way to eliminate it.[36] Furthermore, the evaluation of the *1/f* noise performance during a long-period under high-vacuum conditions has indicated a significant improvement due to the removal of defects and adsorbates from the surface.[43]

Despite the aforementioned advancements in experimental characterisation of *1/f* noise in 2D-FETs, compact modelling developments, appropriate for circuit designs, are still in an embryonic phase. According to bibliography,[33-51] the well-known *ΔN*, *Δμ*, *ΔR* effects generate *1/f* noise in 2D-FETs similarly to MOSFETs[25-32] and graphene FETs (GFETs).[52] *1/f* noise adds as the number of layers decreases[49] and becomes maximum at SL 2D-FETs where *ΔN* effect prevails. *Δμ* dominates in multilayer devices,[39] similarly to GFETs,[51] while it can have a trivial contribution in subthreshold region in SL to FL 2D-FETs[39] whereas *ΔR* has been shown to be higher in SL devices.[51] Most of the 2D-FET *1/f* noise models cited in literature adopt the straightforward $\sim(g_m/I_D)^2$ approximation of *ΔN* mechanism[36-37, 39, 42, 45, 49] which is valid only under homogeneous channel conditions, as mentioned before. A much more intense *ΔN CMF* contribution to *1/f* noise has been recorded in 2D-FETs in comparison with MOSFETs,[27, 30-32] which again has been modelled by applying the simplified $\sim(1+\alpha_C\mu C_{t,b}I_D/g_m)^2(g_m/I_D)^2$ concept,[35, 40-41, 43, 46] where $\alpha_c$ is the Coulomb scattering coefficient (in *V.s/C*), *μ* the low-field mobility of the device (in $cm^2/(V.s)$) and $C_{t,b}$ the top(back)-dielectric capacitance per unit area (in *μF.cm$^{-2}$*). Some other works follow exclusively the *Δμ* model to describe the *1/f* noise experiments' behaviour[33, 38, 47, 50] while in some occasions both *ΔN*, *Δμ* effects are required to successfully interpret the *1/f* noise measurements.[39, 42] More complicated *ΔN*[34] and *Δμ*[47] models are also proposed but they are based on derivations not suitable for circuit design while they also employ linear region approximations.

In the present study, a compact physics-based *1/f* noise model for SL to FL 2D-FETs is for the first time proposed and precisely validated with experiments for a wide range of 2D devices and operating conditions. The model includes explicit derivations for both *ΔN*, *Δμ* effects without approximating a uniform channel which makes the model precise at any $V_{DS}$ region. The decrease of measured *1/f* noise due to velocity saturation (VS) effect, which is critical at high longitudinal electric fields (short channel lengths and/or increased $V_{DS}$), has been accurately modeled in a previous analysis[52, 53] for GFETs, and such a modelling approach is for the first time implemented here after appropriate adjustments according to the 2D-FETs' underlying physics; VS effect has been recorded to be quite intense in 2D-FETs (saturation velocity $u_{sat}$~$5.10^5$-$5.10^6$ *cm/s*).[54-56] An exact *1/f* noise model which functions accurately at every bias condition with one parameter set, firstly requires a qualitative description of the device $I_D$ at any operation regime.[57-62] Such a performance is ensured by a recently proposed physics-, chemical potential ($V_c$)- based *IV* compact model.[60, 63] Secondary effects such as i) the $V_{DS}$ dependence of threshold voltage $V_{TH}$ (Drain Induced Barrier Lowering-DIBL),[62] ii) the $V_{DS}$ dependence of contact resistance $R_c$[61] and iii) the Channel Length Modulation (CLM)[55] effect, are also implemented in the *IV* model here, to allow for better accuracy and flexibility. Notice that the aforementioned *IV* model (and consequently the proposed *1/f* noise one) are reliable for SL to FL 2D-FETs where the thickness of the channel is quite small allowing to consider a uniform $V_c$ along the vertical direction.[59] Additionally, they are valid for both n- and p-type 2D-FETs by changing the polarity of voltages and charges.[63] Afterwards, *ΔN*, *Δμ* *1/f* noise models are derived locally in the channel and then integrated from source to drain to calculate the total PSD for each mechanism under the consideration of uncorrelated local noise sources.[27, 52-53] *ΔR* contribution has also been contained in the proposed *1/f* model through a straightforward expression already used in silicon[27] and graphene devices.[52-53]

As noted previously, different methods have been applied to reduce experimental *1/f* noise in 2D-FETs and thus, in some cases it can be comparable with measured *1/f* noise in advanced emergent technologies as carbon-nanotube-,[33, 38, 46] graphene-[46] and Nanowire-FETs.[38] It is crucial to conduct such testings properly in order to extract reliable conclusions. For example, Hooge parameter $\alpha_H$ is widely used for such comparisons[33, 38, 46] for 2D-FETs with many layers where *Δμ* effect prevails. Under such conditions and if a linear region is assumed, *Δμ* is proportional to $\sim\alpha_H/(WLC_{t,b}V_{GEFF})$[23] resulting in a bias-independent $\alpha_H$ which leads to trustworthy conclusions; *W*, *L* are the width and length of the device, respectively and $V_{GEFF}$ the overdrive gate voltage. The latter $\alpha_H$-based analysis has been applied in comparisons between different multilayer 2D channel materials as in Ref. 47 where $\alpha_H$ has been found quite lower in 15-layer $MoSe_2$ than 15-layer $MoS_2$ FETs for a similar dielectric ($SiO_2$). To the contrary, for SL to FL FETs, dominant *ΔN* effect provides a different

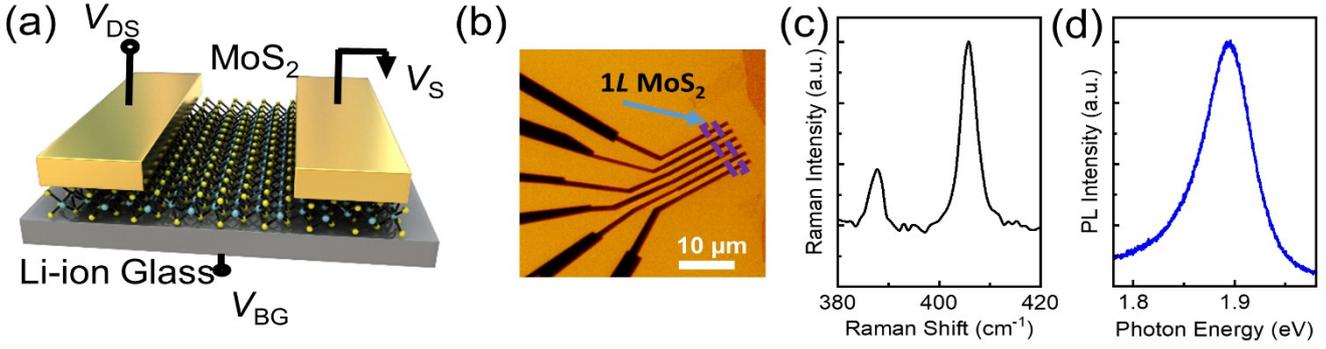

Fig. 1 a) Schematic illustration of a single layer (SL) MoS$_2$ transistor back-gated by Li-ion solid electrolyte substrate.[65] The drain ($V_{DS}$) and back-gate ($V_{GB}$) are biased, and the source is grounded. b) Optical micrograph of SL MoS$_2$ on Li-ion substrate. The purple rectangle marks the boundary of one-layer (1L) MoS$_2$. Scale bar is *10 µm*. c) Raman and d) Photoluminescence (PL) spectrum of a monolayer MoS$_2$ on Li-ion glass substrate.

*1/f* noise bias-dependence than *Δµ*[23, 27-32] and thus, $α_H$ would be bias-dependent if *Δµ* model was exclusively used to describe the specific measurements. The latter must be considered when $α_H$ is used as a comparative tool as different outcomes can be excluded depending on the operation region under examination. In general, since three individual mechanisms [*ΔN* (+*CMF*), *Δµ*, *ΔR*], can affect *1/f* noise, each of which is prominent at different operation regimes, while short channel effects such as VS can also contribute at higher $V_{DS}$ regime, it is preferable to use normalized with area and divided by squared drain current $WL·S_{ID}f/I_D^2$ *1/f* noise vs. $V_{GEFF}$ at linear region for direct comparisons. The aforementioned benchmark is used in this work, for experimental *1/f* noise data taken from bibliography[35, 40, 43, 49] and significant conclusions are derived through straightforward comparisons of the extracted *1/f* noise model parameters between different 2D channel materials and/or dielectrics. Such *1/f* noise experiments are selected carefully to extend from subthreshold to strong accumulation regions in most cases, so as to reveal the contribution of the primary *1/f* noise generators such as *ΔN* effect (accompanied by a strong *CMF* in almost every case) as well as *Δµ*, which is less significant but not negligible especially in deep subthreshold region. Besides, the precise experimental validation of the proposed model for an extensive range of operating conditions and 2D devices, establishes its integrity.

## Devices and Measurement Setup

A back-gated SL MoS$_2$ FET with *W=4 µm* and *L=1 µm* has served as the main DUT after its thorough *IV-1/f* experimental characterization in the present analysis. A Li-ion glass operates both as the back-gate dielectric and the substrate while the SL MoS$_2$ has been CVD grown on SiO$_2$/Si and then has been transferred onto Li-ion glass substrate.[65] The reported effective electrolyte capacitance $C_{EDL}$=2.1 µF/cm$^{-2}$ for the DUT corresponds to a $C_b=ε_0·ε_b/t_b$ with equivalent dielectric thickness $t_b$=1.64 nm,[65] and is used as a parameter in the proposed compact model (where $ε_b$ is the back-dielectric relative permittivity, $ε_0$=8.85·10$^{-12}$ F/m is the absolute permittivity). We have conducted on wafer *IV* and *1/f* noise ($S_{ID}$) measurements at the available DUT[65] for an extensive range of bias conditions, from *1 Hz* to *15 kHz*, so as to comprehensively validate the implemented model and confirm its physical validity. Back-gate voltage $V_{GB}$ and $V_{DS}$ sweeps have been applied from *0 to 1 V* with a step of *100 mV* covering from around $V_{TH}$ to strong accumulation region and from linear to saturation domain; the varied $V_{DS}$ points are essential for the investigation of VS effect on *1/f* noise. A single $S_{ID}f$ value has been extracted for each bias point by averaging for a frequency range from *10 Hz-1 kHz* where *1/f* noise dominates, after applying a filtering technique to remove outliers from the spectra (see Experimental Section for more details on fabrication and measurements procedures). Fig. 1a shows the schematic diagram of Li-ion electrolytic solid substrate device where the mobile ions/vacancies can be accumulated/depleted at the semiconductor channel interface by $V_{GB}$. The accumulated positive ions (negative vacancies) induce electrons (holes) inside the semiconductor channel. Fig. 1b shows an optical image of the fabricated DUT on a SL MoS$_2$ (marked by a purple rectangle) exfoliated on solid electrolyte substrate. Raman and Photoluminescence (PL) spectroscopic images in Fig. 1c-1d show good material quality. Peak separation between in-plane (E$^1_{2g}$) and out-of-plane (A$_{1g}$) Raman modes is approximately *18.3 cm$^{-1}$*, which is typical for a single layer MoS$_2$.[66] The full width at half maximum (FWHM) for E$^1_{2g}$ and A$_{1g}$ Raman modes are *<3 cm$^{-1}$* and *<4 cm$^{-1}$*, respectively, which signifies good crystalline quality of the exfoliated material.[67] The PL peak of the same flake is located at *1.89 eV* and the FWHM is approximately *20 meV*, similar to the reported values in literature for good quality materials.[68]

## Results and Discussion

### Theoretical derivations

For exact *1/f* noise model derivations, an *IV* model, based on drift-diffusion carrier transport with a constant $u_{sat}$ parameter, has been used.[63] The compact $I_D$ expression[63] (Equation 5) is expanded in the present work to include interface trap charge $Q_{it}$ contribution; only $Q_{it}$ effect on device electrostatics, and consequently on $V_c$ estimation, had been considered until now[63] (see Equations S1-S12 and Fig. S1a-b in Section S.1 of the Supporting Information for more details on the definition of different *IV* quantities). In addition, experimental *IV* data from the measured DUT[65], as well as from other 2D-FETs taken from literature,[33, 35, 38, 40, 42-43, 49] have unveiled the urgency for the incorporation of some additional secondary effects (DIBL, $V_{DS}$ dependence of $R_C$, CLM), to the *IV* model to optimize its performance (see Section S.2 of the Supporting Information for more details regarding the implementation of these effects).

The scheme of dividing the channel into microscopic slices and correspondingly, to uncorrelated local noise sources which are integrated from source to drain to derive total PSD for *ΔN*, *Δμ* channel mechanisms, is applied here[27, 52-53] (see Fig. S1c and Equations S13-S15 in Section S.3 of the Supporting Information for more details on the general noise extraction methodology). After applying the appropriate approximations, in order to acquire analytical solutions suitable for circuit simulators (see Equations S16-S26 in Section S.3 of the Supporting Information), the following expressions are derived regarding *ΔN* effect:

$$\Delta N = \Delta N1 + \Delta N2 + \Delta N3 = \Delta N1A - \Delta N1B + \Delta N2 + \Delta N3A - \Delta N3B \quad (1)$$

$$\Delta N1A = \frac{KTN_{ST}e^2}{WL_{eff}C_{dq}U_T^2 C gvc}\left[\frac{\ln(C2+u+C2u)}{(1+C2)}\right]_{u_d}^{u_s} \quad (2)$$

$$\Delta N1B = \frac{KTN_{ST}e^2\mu}{WL_{eff}^2 v_{sat}CC_{dq}U_T}\left|\left[\frac{-1}{(1+C2)(C2+u+C2u)}\right]_{u_s}^{u_d}\right| \quad (3)$$

$$\Delta N2 = \frac{KTN_{ST}L}{WL_{eff}^2}(\alpha_c e\mu)^2 \quad (4)$$

$$\Delta N3A = \frac{2KTN_{ST}e^2\alpha_c\mu}{WL_{eff}CU_T gvc}(u_s - u_d) \quad (5)$$

$$\Delta N3B = \frac{2KTN_{ST}e^2\alpha_c\mu^2}{WL_{eff}^2 v_{sat}C}\left|\left[\frac{\ln(C2+u+C2u)}{(1+C2)}\right]_{u_s}^{u_d}\right| \quad (6)$$

where $L_{eff}$ elucidates an effective channel length which diminishes $I_D$ when VS effect becomes critical (see Equation S8 in Section S.1 of the Supporting Information), while $g_{vc}$ is a normalized current term (see Equation S11 in Section S.1 of the Supporting Information), $u_{s(d)}$ is the source(drain) side normalized charge, calculated through $V_c$[63] (See Section S.1 of the Supporting Information), $U_T$ is the thermal voltage, $e$ is the electron charge, $C_{dq}$ is the degenerated quantum capacitance, $K$ the Boltzmann constant, $T$ the temperature, $C2=(C+C_{it})/C_{dq}$ and $C=C_t+C_b$ with $C_{it}$ the interface trap capacitance[58, 60, 63]. $N_{ST}$ is the slow border trap density per unit energy in $eV^{-1}.cm^{-2}$; $N_{ST}$ and $\alpha_c$ are used as *ΔN 1/f* noise model parameters. *ΔN1* is related to carrier number fluctuation and *ΔN2*, *ΔN3* to CMF effect, respectively. *ΔN1(3)B* represents the *ΔN1(3) 1/f* noise reduction due to VS effect which is also induced through $L_{eff}$ in both *ΔN1(3)A* and *ΔN1(3)B* as well as in *ΔN2* (See Section S.3 of the Supporting Information).

By following an Identical procedure as in *ΔN* case, the following analytical expressions are derived regarding *Δμ* effect (see Equations S27-S29 in Section S.3 of the Supporting Information):

---

[a] Measurements are always symbolized with markers and models with lines throughout the entire manuscript. A detailed *IV* parameter extraction procedure is described in Section S.4 of the Supporting Information, which is applied for every device analysed in the present work. Also note in the right subplot of Fig. 2 that there is a minimum $I_D$ floor (~0.045 $\mu A/\mu m$) in our measurement setup.

$$\Delta\mu = \Delta\mu A - \Delta\mu B \quad (7)$$

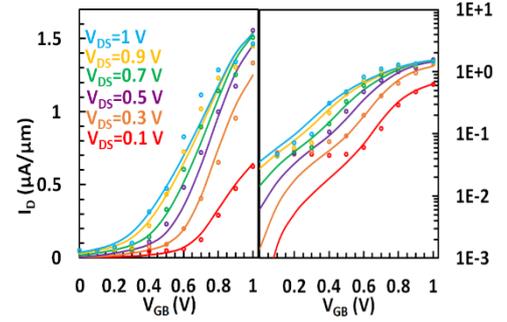

**Fig. 2** Transfer characteristics at $V_{DS}$=0.1, 0.3, 0.5, 0.7, 0.9, 1 V at linear (left subplot) and logarithmic (right subplot) y-axis for Li-ion glass substrate SL MoS$_2$ FET[65] with width *W=4 μm* and length *L=1 μm*. Markers: measurements, lines: model.

$$\Delta\mu A = \frac{\alpha_H e}{WL_{eff}CU_T gvc}\left[u + C2\ln[1-e^u]\right]_{u_d}^{u_s} \quad (8)$$

$$\Delta\mu B = \frac{\alpha_H e\mu}{CWL_{eff}^2 v_{sat}}\left|[\ln[u]]_{u_s}^{u_d}\right| \quad (9)$$

where unitless $\alpha_H$ is used as *Δμ 1/f* noise model parameter. *ΔμB* represents the *Δμ 1/f* noise reduction due to VS effect which is also induced through $L_{eff}$ in both *ΔμA* and *ΔμB* terms. Finally, contact resistance contribution to *1/f* noise (*ΔR*) is provided by a simple approach:[27 (Equation 6.87), 53]

$$\Delta R = \frac{g_{ms}^2 + g_{md}^2}{\left[1+\frac{R_C}{2}(g_{ms}+g_{md})\right]^2}S_{\Delta R^2} \quad (10)$$

where $R_C=R_{CS}+R_{CD}$ is the sum of contact resistances in source and drain side, respectively, extracted from *IV* data, $S_{\Delta R^2}$ in $\Omega^2/Hz$ is the resistance fluctuation used as *ΔR 1/f* noise model parameter and $g_{ms}$, $g_{md}$ are the source and drain transconductances[52-53] (see Equation S30 in Section S.3 of the Supporting Information). Conclusively, total *1/f* noise is given as:

$$\frac{S_{I_D}}{I_D^2}f = \Delta N + \Delta\mu + \Delta R \quad (11)$$

The derived *1/f* noise model is implemented in Verilog-A as a module of the large-signal model proposed in Ref. 63, and can be included in Keysight Advanced Design System (ADS) standard circuit simulator tool.

**Experimental Validation**

Several transport parameters are critical for the *1/f* noise model (cf. Equations 1-11), which must be extracted precisely. Thus, the *IV* model[63] is validated with our DC measurements at the main DUT[65] and the decent consistency between the simulated and measured $I_D$ is depicted in the transfer characteristics shown in Fig. 2 at both linear (left subplot) and logarithmic (right subplot) y-axis, from around $V_{TH}$ to strong accumulation regime, for several $V_{DS}$ values (see Supporting Information: Fig. S3a in Section S.5 for the transfer characteristics for all $V_{DS}$ values, and row 3 of Table S1 in Section S.6 for the derived *IV* parameters); Extracted *IV* parameters are close to the values cited in Ref. 65 with some justified inconsistencies recorded in hysteresis-induced parameters due to dissimilar measurement setups (see Section

S.6 of the Supporting Information).[a] Before proceeding with the bias-dependent *1/f* noise model validation, two measured noise

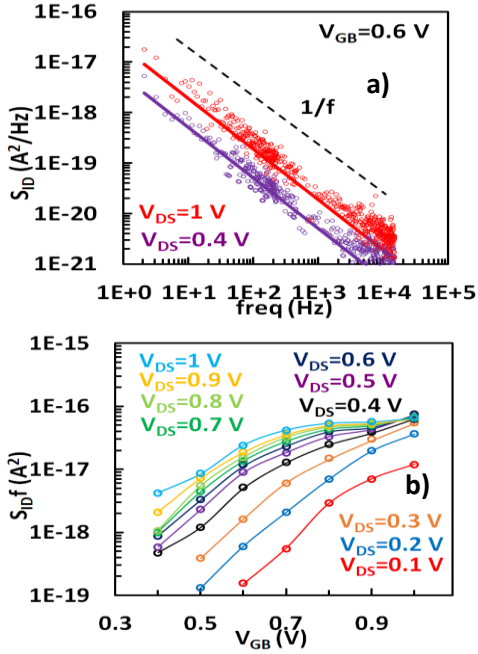

**Fig. 3** a) Drain current noise $S_{ID}$ vs. frequency $f$ for $V_{GB}$=0.6 V at $V_{DS}$= 0.4, 1 V, b) vs. $V_{GB}$ at $V_{DS}$=0.1 to 1 V (0.1 V step), f=1 Hz for Li-ion glass substrate SL MoS$_2$ FET[65] with W=4 µm and L=1 µm. Markers: measurements, solid lines: model. the dashed line in (a) corresponds to a *1/f* slope.

spectra are presented in Fig. 3a for the DUT[65] at $V_{GB}$=0.6 V for a medium (0.4 V) and a high (1 V) $V_{DS}$. The *1/f* trend (dashed line) is apparent in both cases while the model is also shown to accurately fit the experiments. The contribution of thermal noise can become detectable near 1 kHz, for lower bias conditions (see Fig. S3b in Section S.5 of the Supporting Information, where more noise spectra for a wide range of $V_{DS}$, $V_{GS}$ values, are presented). Measured and simulated $S_{ID}f$ vs. $V_{GB}$ at Fig. 3b are demonstrated for all $V_{DS}$ points where the experiments are rigorously interpreted by the model for every case. $S_{ID}f$ increases with both $V_{GB}$ and $V_{DS}$ until it saturates ($S_{ID}f/I_D^2$ is presented to decrease with normalized -divided with W- $I_D$ at Fig. S3c in Section S.5 of the Supporting Information).

$S_{ID}f/I_D^2$ is also depicted vs. normalized $I_D$ in Fig. 4a at $V_{DS}$=0.5 V (left subplot) and at $V_{DS}$=1 V (right subplot) where apart from the measurements and the total model, the contributions from the different *1/f* noise mechanisms are presented with dashed lines. It is apparent that *ΔN* with an intense *CMF* effect, is the dominant mechanism from moderate to strong accumulation region[27] (Section 6.3) for both $V_{DS}$, while the slight increase of the experimental data towards lesser $I_D$ is captured by a soft *Δµ* model which gets more effective deeper in subthreshold regime, similarly to MOSFETs[27] (Section 6.3); due to measurement system limitations, it has not been possible to measure *1/f* noise in extended subthreshold region. A $V_{DS}$-dependent analysis of $S_{ID}f/I_D^2$ has been performed and illustrated in Fig. 4b for lower (0.4, 0.5, 0.6, 0.7 V) and higher (0.8, 0.9, 1 V) $V_{GB}$ in left and right subplots, respectively, where a long-channel model, after de-activating VS effect, is also exhibited with dashed lines for comparison reasons. VS effect contribution moderately reduces *1/f* noise experiments as horizontal electric field elevates and the total model follows this trend in contrary to the long channel approach which overestimates measurements

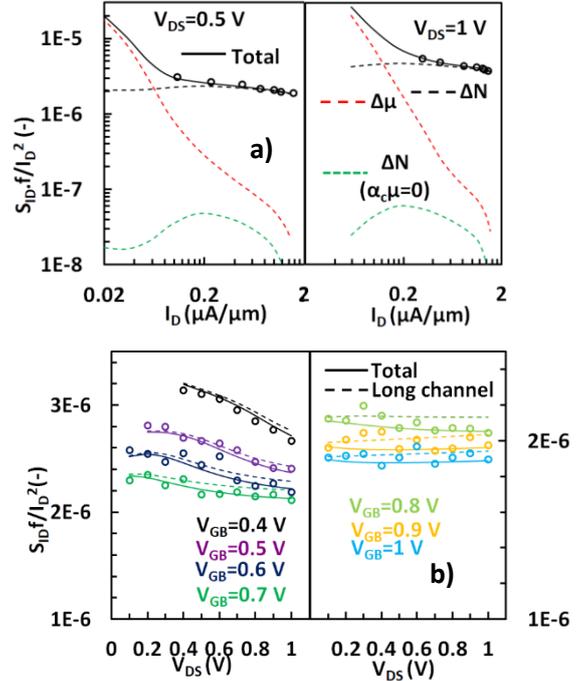

**Fig. 4** a) $S_{ID}f/I_D^2$ vs. normalized drain current $I_D$ at $V_{DS}$=0.5 V (left subplot) and $V_{DS}$=1 V (right subplot), and b) vs. $V_{DS}$ at $V_{GB}$=0.1 to 1 V (0.1 V step) for Li-ion glass substrate SL MoS$_2$ FET[65] with W=4 µm and L=1 µm. Markers: measurements, solid lines: model. dashed lines: Carrier number fluctuation *ΔN* model with (black) and without (green) Coulomb scattering (CMF) effect ($α_cµ$) and mobility fluctuations *Δµ* model (red) (a), long channel model with de-activated velocity saturation (VS) effect (b).

at higher $V_{DS}$ (right subplot). Note that the remarkable behavior of the model for every operation regime is accomplished with one *1/f* noise model parameter set (see row 3 of Table S2 in Section S.6 of the Supporting Information); the extracted *1/f* noise parameters are comparable or even improved than other MoS$_2$ FETs in literature (see Section S.6 of the Supporting Information). Since one of the principal scopes of the present study is to reveal the VS influence on *1/f* noise, a back-gated FL (10-layers) MoSe$_2$ device with a 90 nm SiO2 dielectric and W=2 µm, L=2 µm which presents a more intense VS effect, has been carefully selected from literature[38] in order to highlight the *1/f* noise reduction induced by a relatively low $u_{sat}$ value. Initially, the DC model is perfectly fitted to experimental data for a wide range of operating conditions including subthreshold regime (see Supporting Information: Fig. S4 in Section S.5, and row 4 of Table S1 in Section S.6 for the extracted IV parameters). Experimental and modeled $S_{ID}f/I_D^2$ are demonstrated in Fig. 5a vs. $I_D$ for a low (0.1 V) and a high (0.4 V) $V_{DS}$ value, in left and right subplots, respectively, where the different *1/f* noise mechanisms are introduced with dashed-dotted lines. The model describes flawlessly the measured data while the domination of *ΔN* mechanism with a powerful *CMF* effect, is evident as well as the noticeable contribution of *Δµ* at around $I_D$≈1-3·10$^{-7}$ A for both $V_{DS}$ values in subthreshold region (see transfer characteristics of Ref. 38 at logarithmic y-axis in Fig. S4b in S.5 of the Supporting Information).

A not-negligible addition of $\Delta R$ model can also be observed at very high current regime.

Measurements and total models of $S_{ID}f/I_D^2$ are illustrated vs. $I_D$ in

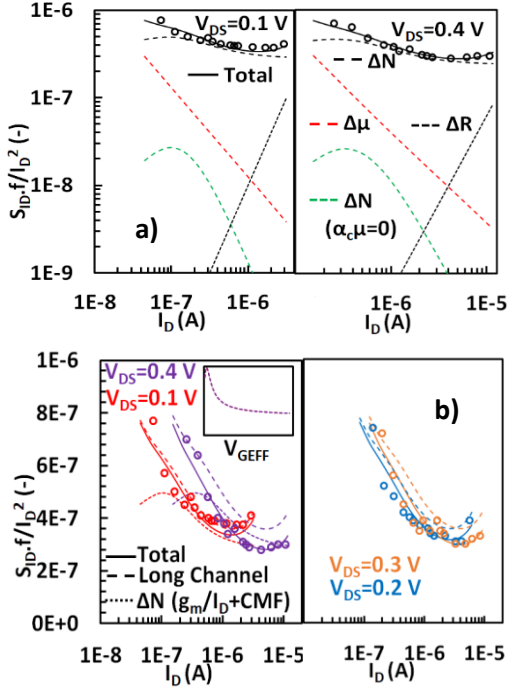

**Fig. 5** $S_{ID}f/I_D^2$ vs. $I_D$ at a) $V_{DS}$=0.1 V (left subplot) and $V_{DS}$=0.4 V (right subplot), and b) $V_{DS}$=0.1 to 0.4 V (0.1 V step) for 10-layer MoSe$_2$ FET[38] with W=2 μm and L=2 μm. Markers: measurements, solid lines: models. Dashed lines: $\Delta N$ model with (black) and without (green) CMF effect ($\alpha_C\mu$) and $\Delta\mu$ model (red) (a), long channel model with de-activated VS effect (b). Dotted lines: Series resistance 1/f noise model $\Delta R$ (a) and simplified $\sim(1+\alpha_C\mu C_{t,b} I_D/g_m)^2 (g_m/I_D)^2$ $\Delta N$ CMF model for $V_{DS}$=0.1, 0.4 V (b); the latter is also shown in the inset of (b) vs back-gate voltage overdrive $V_{GEFF}$ for $V_{DS}$=0.1, 0.4 V.

Fig. 5b for all available $V_{DS}$ where a long-channel model is included in the plots with dashed lines. The total model accurately describes the beneficial reduction of experiments in comparison with the long-channel case as $V_{DS}$ increases, especially at $V_{DS}$=0.4 V, making the necessity of the present analysis prominent (see row 4 of Table S2 in Section S.6 of the Supporting Information for the extracted 1/f noise model parameters). A simple $\sim(1+\alpha_C\mu C_{t,b}I_D/g_m)^2 (g_m/I_D)^2$ $\Delta N$ CMF model[23, 30-32] is also implemented for the $V_{DS}$=0.1, 0.4 V cases, respectively, and depicted in the left subplot of Fig. 5b with dotted lines. The expected agreement of this simple approach with our model is revealed for the low $V_{DS}$ for identical $\Delta N$ model parameters ($N_{ST}$, $\alpha_C$). At the higher $V_{DS}$, the simplified model errs, as no dependence with $V_{DS}$ is recorded regarding its magnitude, just a $V_{TH}$-induced fluctuation towards x-axis. The latter is confirmed in the inset of Fig. 5b where the simple $\Delta N$ CMF model is shown to be the same vs. $V_{GEFF}=V_{GB}-(V_{G(S)B0}-n_{DIBL}V_{DS})-Q_{it}/C$ for both $V_{DS}$ cases examined; the medium term of $V_{GEFF}$ ($n_{DIBL}V_{DS}$) represents the shift of $V_{TH}$ due to DIBL effect (see Section S.2 of the Supporting Information) while the last term ($Q_{it}/C$) depicts a corresponding $V_{TH}$ shift due to interface trapping effects.

In addition, a theoretical investigation is conducted and presented in Fig. 6a where simulated $S_{ID}f/I_D^2$ is demonstrated vs. $V_{GB}$ for back-gated 2D-FETs with i) different SL channel materials (MoS$_2$, WS$_2$, MoSe$_2$ and MoTe$_2$) and the same $C_b$=0.012 μF/cm$^2$

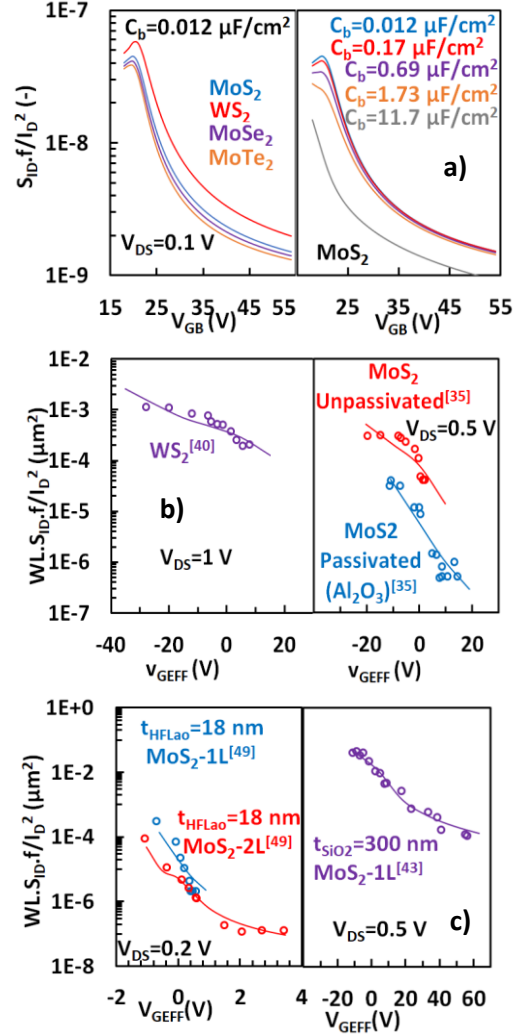

**Fig. 6** a) $S_{ID}f/I_D^2$ vs. $V_{GB}$ simulations for different SL 2D-channel materials (left subplot) and dielectric capacitances (right subplot) for the same IV-1/f noise model parameters. Normalized with area $WL \cdot S_{ID}f/I_D^2$ vs. $V_{GEFF}$ for b) a SL WS$_2$ FET[40] (magenta) with W=27 μm and L=2 μm at $V_{DS}$=1 V (left subplot), Al$_2$O$_3$ passivated (blue) and unpassivated (red) SL MoS$_2$ FET[35] with W=1 μm and L=0.4 μm at $V_{DS}$=0.5 V (right subplot) and for c) bilayer (red) and SL (blue) MoS$_2$ FET[49] with W=15 μm and L=0.3 μm at $V_{DS}$=0.2 V (left subplot) and a SL MoS$_2$ FET[43] exactly after fabricated (magenta) in vacuum system with W=10 μm and L=25 μm at $V_{DS}$=0.5 V (right subplot). Markers: measurements, lines: models.

(left subplot), ii) different $C_b$ and an identical SL MoS$_2$ channel material (right subplot) at low $V_{DS}$; IV parameters from the device depicted in Fig. 5 are employed in both cases.[38] Regarding 1/f noise parameters, $N_{ST}$ is kept the same with the aforestated FET,[38] $\alpha_c$ is downgraded a bit (6·10$^4$ Vs/C) since in case of a severe CMF effect, $\Delta N2$, which is independent of $C_b$ (dielectric thickness) and $C_{dq}$ (2D material) (cf. Equation 4), would prevail and thus, 1/f noise would be identical for every material and dielectric thickness. $\alpha_H$=5·10$^{-5}$ is also reduced to achieve a more balanced contribution of $\Delta N$, $\Delta\mu$ effects, respectively. To simulate 2D-FETs with different 2D materials in scenario (i), the parameters that define 2D density of states $D_{oS}$ (see Section S.1 of the Supporting Information) must be set accordingly[3-5] and conclusively, WS$_2$ is shown to endure the higher *1/f* noise. As the same *1/f* noise parameter set is used for all the devices under discussion, *1/f* noise depends only on $C_{dq}$ (and consequently on $D_{oS}$) as it is displayed in Equations 2, 3, 5, 6, 9. Regarding scenario (ii), notice that $C_b$ is modified only by adjusting the dielectric thickness $t_b$ (and not dielectric material-related $\varepsilon_b$) in the right subplot of Fig. 6a, since if the dielectric material changes (e.g. from SiO$_2$ to Al$_2$O$_3$), the *ΔN*-related parameters, which are strongly related with trapping effects through the dielectric, will probably fluctuate and experimental *1/f* noise data would be essential in order to extract them. Under the above conditions, it is clear that *1/f* noise adds with $t_b$ (or equivalently is inversely proportional to $C_b \sim 1/t_b$) for 2D-FETs with alike dielectrics and channel materials.

It is obvious that direct 1/f noise comparisons of both the model and experiments for 2D-FETs with dissimilar dielectrics or channel materials (after the appropriate normalizations) would offer a better insight regarding the behavior of such devices.[35, 40, 43, 49] Therefore, experimental data from three different fabricated back-gated 2D-FETs with a similar ~*300 nm* SiO$_2$ dielectric, taken from bibliography, are compared next. Hence, unpassivated and top- passivated (with Al2O3) SL MoS$_2$ FETs[35] with *W=1 μm* and *L=0.4 μm* at $V_{DS}$=0.5 V as well as a SL WS$_2$ FET[40] with *W=27 μm* and *L=2 μm* at $V_{DS}$=1 V, are investigated. DC models of these three transistors are calibrated to notably capture the measurements for a wide range of bias conditions (see Supporting Information: Fig. S5 in Section S.5, and rows 5-7 of Table S1 in Section S.6 for the extracted *IV* parameters). Measured and modeled $WL \cdot S_{ID}f/I_D^2$ are demonstrated in Fig. 6b vs. $V_{GEFF}$, with remarkable agreement. $S_{ID}f/I_D^2$ is normalized with area in y-axis to cancel out different device dimensions since *1/f* noise is known to be inversely proportional to device area, while $V_{GEFF}$ is assigned in x-axis in order to ensure comparisons of the devices between equivalent operating regions. After those appropriate normalizations, it is evident that SL WS$_2$ FET introduces higher measured *1/f* noise than SL MoS$_2$ and the model follows this trend. Also, the enhancement of *1/f* noise due to the top-passivation with a high-k dielectric is confirmed (see rows 5-7 of Table S2 in Section S.6 of the Supporting Information for the extracted *1/f* noise model parameters); where WS$_2$ device presents the higher *ΔN1, Δμ 1/f* noise, respectively, with the unpassivated MoS$_2$ to follow and the top-passivated to demonstrate the best performance. In terms of *CMF* contribution, the latter device is still the optimal option (see Section S.6 of the Supporting Information). In the preceding theoretical investigation (cf. Fig. 6a), SL WS$_2$ had evinced the higher *1/f* noise among other 2D materials for the same *1/f* noise parameter set due to the material's intrinsic properties while Fig. 6b experimentally proves that the specific material endures even larger increase in measured *1/f* noise than SL MoS$_2$ probably due to more intense trapping effects, as confirmed from the extracted *1/f* noise parameters.

Subsequently, Fig. 6c depicts the *1/f* noise comparisons between three back-gated MoS$_2$-FETs with different dielectrics. More thoroughly, SL and BL MoS$_2$ FETs[49] with *W=15 μm* and *L=0.3 μm* at $V_{DS}$=0.2 V with an *18 nm* high-k dielectric (HFLaO) in addition to a SL MoS$_2$ FET[43] with *W=10 μm* and *L=25 μm* at $V_{DS}$=0.5 V with a *300 nm* SiO$_2$ dielectric, are examined. A remarkable coherence between the $I_D$ measurements and models is indicated for a wide range of bias points for the aforesaid FETs (see Supporting Information: Fig. S6a-6b, Fig. S7 in Section S.5, and rows 8-10 of Table S1 in Section S.6 for the *IV* extracted parameters). Notice that the experiments of the third device[43] in Fig. 6c (right subplot) are conducted just after the fabrication and placement in a vacuum system.[b] Both $WL \cdot S_{ID}f/I_D^2$ experiments and model are shown in Fig. 6c vs. $V_{GEFF}$ where high-k dielectric FETs are demonstrated in the left subplot while SiO$_2$ one in the right subplot. The former present lower experimental *1/f* noise than the latter which is even lesser in the BL device as it has been expected; the models precisely follow the experiments in every case (see rows 8-10 of Table S2 in Section S.6 of the Supporting Information for the extracted *1/f* noise model parameters); *CMF* is negligible for all of devices under discussion while the high-k dielectric (SL-BL) MoS$_2$ FETs are confirmed to present weaker trapping effects than SiO$_2$ devices (similar extracted values with Refs 43, 49), but slightly more intense Hooge contribution (see Section S.6 of the Supporting Information).

Two more devices[33, 42] are investigated in terms of $I_D$ and *1/f* noise with excellent results regarding the validation of the models with measurements as presented in Fig S8, S9 in Section S.5 of the Supporting Information while the relevant *IV* and *1/f* noise parameters are illustrated in rows 13-15 of Tables S1, S2 in Section S.6 of the Supporting Information, respectively. Regarding the first SL MoS$_2$ FET,[33] *ΔN* effect mainly defines *1/f* noise (see Section S.6 of the Supporting Information). Finally, regarding the FL MoTe$_2$ FETs examined next[42], *1/f* noise in top-passivated devices (HFO$_2$) is once again evidenced to be improved than the unpassivated ones while high $V_{DS}$ is again confirmed to reduce (even slightly) *1/f* noise due to VS effect.

## Conclusions

A thorough physics-based *1/f* noise compact model for SL to FL 2D-FETs has been proposed for the first time, which precisely validates the bias-dependency of experimental *1/f* noise for a wide range of operating conditions with one parameter set. The main DUT investigated in the present work[65] is fully characterized by complete *IV-1/f* noise measurements while additional experimental data are retrieved from literature for FETs with various dielectrics and 2D channel materials. The model functions remarkably for every case also capturing the reduction of measured *1/f* noise, induced by the VS effect at high longitudinal electric fields at shorter gate lengths and/or heightened $V_{DS}$, contrary to the existing $\sim(g_m/I_D)^2$-related $\Delta N$ (+CMF) simplified approaches taken from CMOS, that are valid only under uniform channel conditions. Given the significance of employing scaled-down devices for achieving frequencies at *GHz* range suitable for RF applications, the accurate prediction of VS effect on *1/f* noise experiments is critical. The suggested model is implemented in Verilog-A and can be incorporated in circuit simulators which is a beneficial advance towards the deployment of 2D-circuit design and fabrication.

---

[b] *IV-1/f* noise measurements for the SL MoS$_2$ FET[43], are also conducted after 2 weeks and 8 months (see Supporting Information: Fig. S6 in Section S.5, and rows 11-12 of Tables S1, S2 in Section S.6 for the extracted *IV-1/f* model parameters), where the more the device has remained in the vacuum before measurements, the more the $I_D$ increases and $S_{ID}f/I_D^2$ diminishes while the models accurately adhere to this behaviour.

$\Delta N$ effect due to trapping/detrapping mainly defines *1/f* noise in SL to FL 2D-FETs while the influence of trapping on Coulomb scattering effect has been verified to be significant in the specific transistors, especially from medium to high current regime. $\Delta\mu$ mechanism, which is known to prevail in multilayer devices, is also demonstrated to establish a slight increase of *1/f* noise at subthreshold region for SL to FL devices. In some cases, $\Delta R$ adds to total *1/f* noise at very high current regime. The above mechanisms are included in the compact model where, despite following an equivalent methodology as in GFETs[52], critical modifications in the formulations of the model are induced by the different physical mechanisms that govern specifically 2D-FETs, which have not been considered in any prior work.

The plethora of available 2D materials and dielectrics in addition with the still premature phase of the technology under discussion, has led to many comparisons between different devices in terms of *1/f* noise. Such comparisons have been mainly conducted based on $\alpha_H$ parameter so far in literature but for SL to FL devices where several mechanisms can contribute to *1/f* noise, a more reliable benchmark is applied in the present study where $WL \cdot S_{ID}/I_D^2$ *1/f* noise is mainly investigated. More specifically, to analyse different 2D channel materials, 2D-FETs with the same dielectric (type and thickness) are employed in order to counteract any effect from the dielectric on *1/f* noise and hence, to ensure valid conclusions regarding the 2D materials' *1/f* noise performance. Similarly, when different types of dielectrics are to be examined, devices with identical channel material are chosen. $WS_2$ is theoretically and experimentally evidenced to present higher *1/f* noise than other 2D materials while high-k dielectric materials are confirmed to improve *1/f* noise. For transistors with similar channel material and dielectric, simulations predict a decrease of *1/f* noise as dielectric gets thinner. To conclude, the physics based compact model proposed in this work, which equivalently considers all the physical mechanisms that define *1/f* noise in 2D-FETs, allows for trustworthy outcomes regarding the behaviour of *1/f* noise in these devices.

## Conflicts of interest

There are no conflicts to declare

## Author contributions

**Nikolaos Mavredakis**: Conceptualization, Formal Analysis, Methodology, Software, Validation, Writing-Original draft. **Nikolaos Mavredakis, Anibal Pacheco-Sanchez, Md Hasibul Alam**: Data curation, Visualization. **Md Hasibul Alam**: Resources. **Nikolaos Mavredakis, Anibal Pacheco-Sanchez, Anton Guimerà-Brunetc, Javier Martinez**: Investigation. **Jose Antonio Garrido, Deji Akinwande, David Jiménez**: Supervision. **Nikolaos Mavredakis, Anibal Pacheco-Sanchez, Md Hasibul Alam**: Writing- Reviewing and Editing.

## Experimental Section

### Fabrication of Li-ion MoS$_2$ FETs

Lithium-Ion Conductive Glass Ceramic (LICGC$^{TM}$) AG-01 from Ohara Corporation was used as Li-ion solid electrolyte in this work. $MoS_2$ is exfoliated from bulk crystal (purchased from commercial vendor 2D Semiconductors) using ultra tape (Ultra Tape 1310) and transferred onto Li-ion glass substrates from ultra-tape using polydimethylsiloxane (PDMS) stamp. The exfoliated material was patterned with electron beam lithography (EBL) and $Cl_2/O_2$ plasma etching was used to define channel region. Next, EBL was used to pattern drain/source contact, followed by e-beam deposition ($10^{-6}$ *Torr*) of contact metals (Ni/Au *20 nm/30 nm*) and subsequent lift-off.

### Electrical characterization

The device is experimentally characterized as follows in order to obtain the drain current ($I_D$) and the Low-frequency noise (LFN) spectra: A first pre-amplification stage with a gain of $10^4$ is used to increase the measured $I_D$ signal. In a next high pass filtering stage, low frequency components out of interest for this work (around the DC level) are removed from the signal. The signal is then amplified in a second amplification stage with a gain of $10^4$ followed by an anti-aliasing filtering. An NI DAQ Card is used then to acquire the data with the following procedure. Each bias point is measured, at a specific bias point, after a steady-state current has been found using a sophisticated staircase sweep in which the current is read in frames of *1 s*. The steady-state current is saved in each frame after the slope of the curve ($dI_D/dt$) is below certain error (around $1 \cdot 10^{-8}$ *A/s*). The power spectral density is calculated by acquiring $I_D$ during *1* minute at $50 \cdot 10^3$ *samples/s*.

### Data availability

The data that support the findings of this study are available upon request from Nikolaos Mavredakis (nikolaos.mavredakis@uab.cat). Verilog-A code will be available as soon as it is ready for release.

## Acknowledgements

This work was funded by the European Union's Horizon 2020 Research and innovation Program under Grant Agreement No. GrapheneCore3 881603, Marie Skłodowska-Curie Grant Agreement No 665919 and Grant Agreement No. 732032 (BrainCom). We also acknowledge financial support by Spanish government under the projects RTI2018-097876-B-C21 (MCIU/AEI/FEDER, UE), PID2021-127840NB-I00 and FJC2020-046213-I. The work has also received partial funding from the European Union Regional Development Fund within the framework of the ERDF Operational Program of Catalonia 2014-2020 with the support of the Department de Recerca i Universitat, with a grant of 50% of total cost eligible. GraphCAT project reference: 001-P-001702. The ICN2 is also supported by the Severo Ochoa Centres of Excellence programme, funded by the Spanish Research Agency (AEI, grant no. SEV-2017-0706). We would like to thank Xintong Li from UT Austin for his help with sample preparation.

## References

(1) K. S. Novoselov, A. K. Geim, S. V. Mozorov, Y. Zhang, S. V. Dubonos, I. V. Grigorieva, A. A. Firsov, *Science*, 2004, **306**, 666.

(2) M. Saeed, P. Palacios, M. D. Wei, E. Baskent, C. Y. Fan, B. Uzlu, K. T. Wang, A. Hemmeter, Z. Wang, D. Neumaier, M. C. Lemme, R. Negra, *Adv. Mater.*, 2022, **34**, 2108473.

(3) V. Mishra, S. Smith, K. Ganapathi, S. Salahuddin, *IEEE International Electron Devices Meeting*, 2013, 5.6.1.

(4) J. Chang, L. F. Register, S. K. Banerjee, *Appl. Phys. Lett.*, 2014, **115**, 084506.

(5) Y. Sun, X. Wang, X. G. Zhao, Z. Shi, L. Zhang, *J. Semicond.*, 2018, **39**, 072001.

(6) G. R. Bhimanapati, Z. Lin, V. Meunier, Y. Jung, J. Cha, S. Das, D. Xiao, Y. Son, M. S. Strano, V. R. Cooper, L. Liang, S. G. Louie, E. Ringe, W. Zhou, S. S. Kim, R. R. Naik, B. G. Sumpter, H. Terrones, F. Xia, Y. Wang, J. Zhu, D. Akinwande, N. Alem, J. A. Schuller, R. E. Schaak, M. Terrones, J. A. Robinson, *ACS Nano*, 2011, **9**, 11509.

(7) D. Akinwande, C. Huyghebaert, C. H. Wang, M. I. Serna, S. Goossens, L.-J. Li, H. S. Philip Wong, F. H. L. Koppens, *Nature*, 2019, **573**, 507.

(8) B. Radisavljevic, A. Radenovic, J. Brivio, V. Giacometti, A. Kis, *Nature Nanotech*, 2011, **6**, 147.

(9) S. Larentis, B. Fallahazad, E. Tutuc, *Appl. Phys. Lett.*, 2012, **101**, 223104.

(10) F. Withers, T. H. Bointon, D. C. Hudson, M. F. Craciun, S. Russo, *Sci Rep*, 2014, **4**, 4967.

(11) N. R. Pradhan, D. Rhodes, S. Feng, Y. Xin, S. Memaran, B. H. Moon, H. Terrones, M. Terrones, L. Balicas, *ACS Nano*, 2014, **8**, 5911.

(12) C. M. Corbet, C. McClellan, A. Rai, S. S. Sonde, E. Tutuc, S. K. Banerjee, *ACS Nano*, 2015, **9**, 363.

(13) L. Li, Y. Yu, G. J. Ye, Q. Ge, X. Ou, H. Wu, D. Feng, X. H. Chen, Y. Zhang, *Nature Nanotech*, 2014, **9**, 372.

(14) B. Radisavljevic, M. B. Whitwick, A. Kis, *ACS Nano*, 2011, **5**, 9934.

(15) H. Wang, L. Yu, Y. H. Lee, Y. Shi, A. Hsu, M. L. Chin, L. J. Li, M. Dubey, J. Kong, T. Palacios, *Nano Lett.*, 2012, **12**, 4674.

(16) S. Das, H. Y. Chen, A. V. Penumatcha, J. Appenzeller, *Nano Lett.*, 2013, **13**, 100.

(17) G. Eda, H. Yamaguchi, D. Voiry, T. Fujita, M. Chen, M. Chhowalla, *Nano Lett.*, 2011, **11**, 5111.

(18) O. Lopez-Sanchez, D. Lembke, M. Kayci, A. Radenovic, A. Kis, *Nature Nanotech*, 2013, **8**, 497.

(19) F. Xia, H. Wang, D. Xiao, M. Dubey, A. Ramasubramaniam, *Nature Photon*, 2014, **8**, 899.

(20) H. Li, Z. Yin, Q. He, H. Li, X. Huang, G. Lu, D. W. H. Fam, A. L. Y. Tok, Q. Zhang, H. Zhang, *Small*, 2011, **8**, 63.

(21) F. K. Perkins, A. L. Friedman, E. Cobas, P. M. Campbell, G. G. Jernigan, B. T. Jonker, *Nano Lett.*, 2013, **13**, 668.

(22) D. Sarkar, W. Liu, X. Xie, A. C. Anselmo, S. Mitragotri, K. Banerjee, *ACS Nano*, 2014, **8**, 3992.

(23) G. Ghibaudo, T. Boutchacha, *Microelectron Reliab*, 2002, **42**, 573.

(24) S. K. Lee, C. G. Kang, Y. G. Lee, C. Cho, E. Park, H. J. Chung, S. Seo, H. D. Lee, B. H. Lee, *Carbon*, 2012, **50**, 4046.

(25) M. J. Kirton, M. J. Uren, *Advances in Physics*, 1989, **38**, 367.

(26) F. N. Hooge, *Physica B+C*, 1976, **83**, 14-23.

(27) C. Enz, E. Vitoz, *Charged-based MOS Transistor Modeling*, John Wiley and Sons, Hoboken, NJ, USA 2006.

(28) G. Reimbold, *IEEE Trans Electron Devices*, 1984, **31**, 1190-1194.

(29) G. Ghibaudo, *Solid-State Electron*, 1989, **32**, 563-565.

(30) K. K. Hung, P. K. Ko, C. Hu and Y. C. Cheng, *IEEE Trans Electron Devices*, 1990, **37**, 654-665.

(31) G. Ghibaudo, O. Roux, Ch. Nguyen-Duc, F. Balestra and J. Brini, *Phys. Stat. sol. (a)*, 1991, **124**, 571.

(32) T. Boutchacha, G. Ghibaudo, *IEEE Trans Electron Devices*, 2011, **58**, 3156-3161.

(33) V. K. Sangwan, H. N. Arnold, D. Jariwala, T. J. Marks, L. J. Lauhon, M. C. Hersam, *Nano Lett.*, 2013, **13**, 4351.

(34) X. Xie, D. Sarkar, W. Liu, J. Kang, O. Marinov, M. J. Deen, K. Banerjee, *ACS Nano*, 2014, **8**, 5633.

(35) D. Sharma, M. Amani, A. Motayed, P. B. Shah, A. G. Birdwell, S. Najmaei, P. M. Ajayan, J. Lou, M. Dubey, *Nanotechnology*, 2014, **25**, 155702.

(36) S. Ghatak, S. Mukherjee, M. Jain, D. D. Sarma, A. Ghosh, *APL Mater.*, 2014, **2**, 092515.

(37) J. Renteria, R. Samnakay, S. L. Rumyantsev, C. Jiang, P. Goli, M. S. Shur, A. A. Balandin, *Appl. Phys. Lett.*, 2014, **104**, 153104.

(38) S. R. Das, J. Kwon, A. Prakash, C. J. Delker, S. Das, D. B. Janes, *Appl. Phys. Lett.* 2015, **106**, 083507.

(39) D. Sharma, A. Motayed, P. B. Shah, M. Amani, M. Georgieva, A. G. Birdwell, M. Dubey, Q. Li, A. V. Davydov, *Appl. Phys. Lett.*, 2015, **107**, 162102.

(40) M. K. Joo, Y. Yun, S. Yun, Y. H. Lee, D. Suh, *Appl. Phys. Lett.*, 2016, **109**, 153102.

(41) H. Ji, M. K. Joo, Y. Yun, J. H. Park, G. Lee, B. H. Moon, H. Yi, D. Suh, S. C. Lim, *ACS Appl. Mater. Interfaces*, 2016, **8**, 19092.
(42) M. K. Joo, Y. Yun, H. Ji, D. Suh, *Nanotechnology*, 2018, **30**, 035206.
(43) H. Ji, H. Yi, J. Seok, H. Kim, Y. H. Lee, S. C. Lim, *Nanoscale*, 2018, **10**, 10856.
(44) X. Xiong, X. Li, M. Huang, T. Li, T. Gao, Y. Wu, *IEEE Electron Device Lett.*, 2018, **39**, 127.
(45) W. Liao, W. Wei, Y. Tong, W. K. Chim, C. Zhu, *ACS Appl. Mater. Interfaces*, 2018, **10**, 7248.
(46) J. W. Wang, Y. P. Liu, P. H. Chen, M. H. Chuang, A. Pezeshki, D. C. Ling, J. C. Chen, Y. F. Chen, Y. H. Lee, *Adv. Electron. Mater.*, 2018, **4**, 1700340.
(47) J. Kwon, C. J. Delker, C. T. Harris, S. R. Das, D. B. Janes, *J. Appl. Phys.*, 2020, **128**, 094501.
(48) W. Liao, L. Huang, *Phys. Status Solidi RRL*, 2021, **15**, 2100276.
(49) Q. Gao, C. Zhang, Z. Yi, Z. Pan, F. Chi, L. Liu, X. Li, Y. Wu, *Appl. Phys. Lett.*, 2021, **118**, 153103.
(50) L. Pimpolari, G. Calabrese, S. Conti, R. Worsley, S. Majee, D. K. Polyushkin, M. Paur, C. Casiraghi, T. Mueller, G. Iannaccone, M. Macucci, G. Fiori, *Adv. Electron. Mater.*, 2021, **7**, 2100283.
(51) S. L. Rumyantsev, C. Jiang, R. Samnakay, M. S. Shur, A. A. Balandin, *IEEE Electron Device Lett.*, 2015, **36**, 517-519.
(52) F. Pasadas, P. C. Feijoo, N. Mavredakis, A. Pacheco-Sanchez, F. A. Chaves, D. Jiménez, *Adv. Mater.*, 2022, **34**, 2201691.
(53) N. Mavredakis, W. Wei, E. Pallecchi, D. Vignaud, H. Happy, R. G. Cortadella, N. Schaefer, A. B. Calia, J. A. Garrido, D. Jimenez, *IEEE Trans Electron Devices*, 2020, **67**, 2093.
(54) G. He, K. Ghosh, U. Singisetti, H. Ramamoorthy, R. Somphonsane, G. Bohra, M. Matsunaga, A. Higuchi, N. Aoki, S. Najmaei, Y. Gong, X. Zhang, R. Vajtai, P. M. Ajayan, J. P. Bird, *Nano Lett.*, 2015, **15**, 5052.
(55) K. K. H. Smithe, C. D. English, S. V. Suryavanshi, E. Pop, *Nano Lett.*, 2018, **18**, 4516.
(56) A. Sebastian, R. Pendurthi, T. H. Choudhury, J. M. Redwing, S. Das, *Nat Commun*, 2021, **12**, 693.
(57) D. Jimenez, *Applied Physics Letters*, 2012, **101**, 243501.
(58) S. V. Suryavanshi, E. Pop, *J. Appl. Phys.*, 2016, **120**, 224503.
(59) C. Jiang, M. Si, R. Liang, J. Xu, P. D. Ye, M. A. Alam, *IEEE J. Electron Devices Soc.*, 2018, **6**, 189.
(60) E. G. Marin, S. J. Bader, D. Jena, *IEEE Trans Electron Devices*, 2018, **65**, 1239.
(61) J. Cao, S. Peng, W. Liu, Q. Wu, L. Li, D. Geng, G. Yang, Z. Ji, N. Lu, M. Liu, *J. Appl. Phys.*, 2018, **123**, 064501.
(62) C. Yadav, P. Rastogi, T. Zimmer, Y. S. Chauhan, *IEEE Trans Electron Devices*, 2018, **65**, 4202.
(63) F. Pasadas, E. G. Marin, A. Toral-Lopez, F. G. Ruiz, A. Godoy, S. Park, D. Akinwande, D. Jiménez, *npj 2D Mater Appl*, 2019, **3**, 47.
(64) Y. Y. Illarionov, G. Rzepa, M. Waltl, T. Knobloch, A. Grill, M. M. Furchi, T. Mueller, T. Grasser, *2d Mater.*, 2019, **3**, 035004.
(65) M. H. Alam, Z. Xu, S. Chowdhury, Z. Jiang, D. Taneja, S. K. Banerjee, K. Lai, M. H. Braga, D. Akinwande, *Nat Commun*, 2020, **11**, 3203.
(66) H. Li, Q. Zhang, C. C. R. Yap, B. K. Tay, T. H. T. Edwin, A. Olivier, D. Baillargeat, *Advanced Functional Materials*, 2012, **22**, 1385–1390.
(67) M. R. Laskar, L. Ma, S. Kannappan, P. Sung Park, S. Krishnamoorthy, D. N. Nath, W. Lu, Y. Wu, S. Rajan, *Appl. Phys. Lett.*, 2013, **102**, 252108.
(68) A. Splendiani, L. Sun, Y. Zhang, T. Li, J. Kim, C.-Y. Chim, G. Galli, F. Wang, *Nano Lett.*, 2010, **10**, 1271–1275.